\documentclass[onecolumn,showpacs,preprintnumbers,floatfix,amsmath,amssymb,aps]{revtex4}
\usepackage{bm,tikz}
\usepackage{color}

\usepackage{graphicx,subfig,epsfig,epsf,color}
\usepackage[utf8x]{inputenc}
\usepackage{amssymb,amsmath,appendix}
\usepackage{slashed}
\usepackage{array}
\usepackage{hyperref}
\usepackage{ulem}

\begin{document}
\title{Influence of the effective mass on the properties of nuclear matter at finite density and temperature}

\author{Hajime Togashi $^{1, 2}$, Debashree Sen $^{3}$, Hana Gil $^{3}$, and Chang Ho Hyun $^{1,}$*}

\affiliation{$^1$Department of Physics Education, Daegu University, Gyeongsan 38453, Korea\\
$^{2}$Research Center for Nuclear Physics, Osaka University, Osaka 567-0047, Japan}
\affiliation{$^{3}$Center for Extreme Nuclear Matters, Korea University, Seoul 02841, Korea}

\email{togashi2014@gmail.com, debashreesen88@gmail.com, khn1219@gmail.com,  hch@daegu.ac.kr}

\date{\today}




\begin{abstract}

Significance of the chiral symmetry restoration is studied by considering the
role of the modification of the nucleon mass in nuclear medium at finite density
and temperature.
Using the Korea-IBS-Daegu-SKKU density functional theory, we can create models
that have an identical nuclear matter equation of state but different isoscalar 
and isovector effective masses at zero temperature.
Effect of the effective mass becomes transparent at non-zero temperatures,
and it becomes more important as temperature increases.
Role of the effective mass is examined thoroughly by calculating the dependence
of thermodynamic variables such as free energy, internal energy, entropy, pressure
and chemical potential on density, temperature and proton fraction.
We find that sensitivity to the isoscalar effective mass is several times larger than
that of the isovector effective mass, so the uncertainties arising from the effective
mass are dominated by the isoscalar effective mass.
In the analysis of the relative uncertainty, we obtain that the maximum uncertainty
is less than 2 \% for free energy, internal energy and chemical potential, but it
amounts to 20 \% for pressure.
Entropy shows a behavior completely different from the other four variables that
the uncertainty is about 40 \% at the saturation density and increases monotonically
as density increases.
Effect of the uncertainty to properties of physical systems is investigated with
the proto-neutron star.
It is shown that temperature depends strongly on the effective mass at a given density and substantial swelling of the radius occurs due to the finite temperature.
Equation of state is stiffer with smaller isoscalar effective mass,
so the effect of the effective mass appears clearly in the mass-radius relation
of the proto-neutron star, larger radius corresponding to smaller effective mass.\\

{\textbf{Keywords:}} nuclear matter; effective mass; finite temperature; proto-neutron star
\end{abstract}




\maketitle



\section{Introduction}
In the last decade, the {\it ab initio} calculation has become a potential method in the traditional
nuclear physics such as the nuclear structure and reactions.
It provides a way to describe the nuclear interactions in nuclear many-body systems
in terms of the two-, three- and multi-nucleon interactions.
By using the chiral effective field theory (EFT), the inter-nucleon interactions can be expanded in a perturbative way,
so it allows a systematic control of the uncertainties pertinent to nuclear forces and their consequential effects \cite{ceft2013, qmc2015}. 

Chiral EFT is an effective theory of quantum chromo dynamics (QCD) scaled down to low-energy nuclear phenomena.
An essential ingredient that connects the chiral EFT to QCD is the chiral symmetry.
In the chiral EFT, chiral symmetry is broken by the presence of nucleon mass.
From various theoretical calculations, it is shown that the nucleon mass decreases at finite density, so the chiral symmetry 
is restored partially in nuclear medium. 
When the density is high enough that the quarks inside the nucleon become unbound,
it is believed that the chiral symmetry is completely restored.
In-medium mass, more frequently referred to as the effective mass of the nucleon is one of the key uncertainties
that have deep impart to diverse nuclear systems and phenomena \cite{Li:2018lpy, aps2021, prc2023, aps2025}.

Development of the KIDS (Korea-IBS-Daegu-SKKU) nuclear density functional theory is motivated to put the nuclear many-body theory on a systematic expansion scheme on one hand, and on the other hand to reduce the uncertainties pertinent to the nuclear properties and the density dependence of the nuclear matter equation of state (EoS) at both below and above the saturation density ($n_0$). 
A large number of intensive studies in recent years have
constrained the density dependence of the nuclear symmetry energy by using the modern observation data of the neutron star \cite{Li:2021thg, Miyatsu:2023lki, Lim2024, Burgio2024}. 
Thousands of models for the nuclear matter and finite nuclei have been created by using 
various theories, and those consistent with the neutron star data have been sorted out. As a result, the range of the slope ($L$) and curvature ($K_{\rm sym}$) of the symmetry energy has been significantly constrained. 
Similar studies have also been conducted using the KIDS framework, working to further refine the models \cite{prc103, npsm71, ijmpe31}.

Another big issue in the nuclear many-body problem is the value of the isoscalar and isovector effective masses at $n_0$. Conventionally they are constrained in terms of the dynamic properties of nuclei such as various resonance behavior, or in terms of specific conditions for the nuclear matter. 
Empirical ranges of the isoscalar and isovector effective masses are
$(0.7-1.0)m_N$ and $(0.6-0.9)m_N$, respectively (for a recent extensive review,
see \cite{Li:2018lpy} and references therein).
$m_N$ denotes the nucleon mass in free space.
In the KIDS framework, by construction, the nuclear matter EoS is independent of the effective mass at zero temperature \cite{prc99}. However, the effective mass is explicitly included in the thermal fluctuation of the baryon distribution, so their effect is likely to appear directly in the thermodynamic properties of nuclear matter. 
In case of hot nuclear matter and proto-neutron star environment, 
the compositional, isospin asymmetry, and thermal effects largely affect 
the EoS \cite{Prakash:1996xs,Dexheimer:2008ax,Burgio:2020fom,Sedrakian:2021qjw,Sedrakian:2022kgj,Raduta:2020fdn}. Several studies, including numerical simulations \cite{Constantinou:2014hha,Constantinou:2015mna,KN2019,Schneider:2019shi,Yasin:2018ckc,Raithel:2021hye,Andersen:2021vzo,Raduta:2021coc} 
suggest that properties of hot nuclear matter and the corresponding related 
phenomena of supernovae, proto-neutron star, and binary neutron star mergers 
are predominantly affected by the nucleon effective mass. This is because the 
effective mass contributes substantially to the determination of the kinetic 
energy and consequently the strength of thermal effects of the system.

In the present work, constraining the zero-temperature EoS within the range of experimental data or ab initio calculation result, isoscalar and isovector effective masses 
at $n_0$ are fixed to the values 
$(\mu_{IS}, \mu_{IV}) \equiv (m^*_{IS}, m^*_{IV})/m_{N} =$ (0.7, 0.7), (0.7, 0.8), (0.7, 0.9),  (0.7, 1.0),  (0.8, 0.7), (0.8, 0.8), (0.8, 0.9),  (0.8, 1.0), (0.9, 1.0), and (1.0, 1.0),
where $m^*_{IS}$ is isoscalar effective mass, $m^*_{IV}$ is the isovector effective mass, both of which will be defined later.
Ten models that have distinct splitting of the effective mass are applied to
the calculation of the thermodynamic properties of the nuclear matter. 
The thermodynamic variables such as free energy, internal energy, entropy and 
pressure are obtained as functions of density, temperature, and proton fraction. 
In fact, these thermodynamic quantities are essential inputs for numerical simulations of core-collapse supernovae and  the subsequent evolution of proto-neutron stars.
As an application, we calculate the bulk properties of the proto-neutron star for the neutrino-trapped case in the isentropic condition by solving the Tolman-Oppenheimer-Volkoff equations \cite{Tolman:1939jz,Oppenheimer:1939ne}. We find that the thermodynamic variables are sensitive to the effective masses, so the results are distinguished manifestly. The mass-radius relation of the proto-neutron star also depends strongly on the effective masses.

We organize the work in the following order. In Sec. \ref{Sec:model}, we present detailed information for the models, and explain how the finite temperature effect is incorporated in the nuclear matter EoS. 
Section \ref{Sec:result} shows the result for the thermodynamic quantities 
of nuclear matter and discuss the effects of
the uncertainty of the effective masses on the EoS. In Sec. \ref{Sec:PNS}, 
we apply the obtained EoSs to the calculations of proto-neutron star structure. The work is summarized in Sec. \ref{Sec:Sum}.

\section{Model}
\label{Sec:model}

In the KIDS framework, undetermined constants of the functional are determined stepwise. In the first step, we consider the energy per particle in homogeneous nuclear matter 
at temperature $T$ = 0 as
\begin{equation}
E (n_B, Y_p) = \sum_{n=0}^2 n_B^{1+n/3} \alpha_n + 
(1-2Y_p)^2 \sum_{n=0}^3 n_B^{1+n/3} \beta_n,
\label{eq:kids}
\end{equation}
%
%
%
where $n_B=n_n + n_p$ is the baryon density, 
$Y_p = n_p/n_B$ is the proton fraction, and $n_n$ and $n_p$ are densities of the neutron and the proton, respectively. It is shown that three $\alpha$'s ($\alpha_0$, $\alpha_1$, $\alpha_2$) are optimal for a correct description of the symmetric nuclear matter, and it is four for $\beta$ ($\beta_0$, $\beta_1$, $\beta_2$, $\beta_3$) for the asymmetric nuclear matter \cite{kids-nm}. 
In the KIDS0 model, 
which is an initial version of the KIDS density functional, three $\alpha$'s are adjusted to $n_0=$ 0.16 fm$^{-1}$, $E_B = 16.0$ MeV and $K_0=$ 240 MeV, which correspond to the saturation density, binding energy per particle, and incompressibility of the symmetric matter, respectively. Four $\beta$'s are fitted to the pure neutron matter EoS obtained by the variational chain summation method with modern nucleon-nucleon interactions \cite{apr}. 
It is verified that though KIDS0 model is determined independent of the neutron star data, the model satisfies the most recent data of large mass ($M_{\rm max} \geq 2 M_\odot$) and radius of the canonical star ($R_{1.4} =$ 11.8-13.1 km) \cite{miller2021}.

In the second step of the parameter fitting, the model is applied to finite nuclei. 
The least number of additional parameters is two, one is for the density-gradient term, 
and the other is for the spin-orbit interaction \cite{acta}. 
When only two parameters are fitted to nuclear data, there is no restriction to the effective masses, and they are obtained as results of the fitting. 
If we assume specific values for the effective masses, two terms corresponding
to the exchange terms in the density-gradient interaction are added,
and four parameters are adjusted to nuclear data and effective mass.
Skyrme-type force in the KIDS model is given by
\begin{eqnarray}
V_{ij} &=& (t_0 + y_0 P_\sigma) \delta ({\bf r}_{ij})
+ \frac{1}{2} (t_1 + y_1 P_\sigma) [\delta ({\bf r}_{ij}) {\bf k}^2 + {\bf k}^2 \delta ({\bf r}_{ij})] \nonumber \\
&& + (t_2 + y_2 P_\sigma) {\bf k}' \cdot \delta ({\bf r}_{ij}) {\bf k}
+ i W_0 {\bf k}' \times \delta ({\bf r}_{ij}) {\bf k} \cdot
(\vec{\sigma}_i + \vec{\sigma}_j) \nonumber \\
&& + \frac{1}{6} \sum_{n=1}^3 (t_{3n} + y_{3n} P_\sigma)
\rho^{n/3} \delta ({\bf r}_{ij}),
\label{eq:skyrme}
\end{eqnarray}
where ${\bf r}_{ij} = {\bf r}_i - {\bf r}_j$,
${\bf k}=(\nabla_i-\nabla_j)/(2i)$, 
${\bf k}' = -(\nabla'_i - \nabla'_j)/(2i)$,
and $P_\sigma$ is the spin exchange operator.

By using the Skyrme parameters in Eq. (\ref{eq:skyrme}), the energy per particle for homogeneous matter at zero temperature is expressed as 
\begin{eqnarray}
E (n_B, Y_p) &=& \frac{3}{5}(3\pi^2)^{2/3} \left[ Y_p^{5/3} \frac{\hbar^2}{2m_p^*} + (1-Y_p)^{5/3} \frac{\hbar^2}{2m_n^*}\right] n_B^{2/3} \nonumber \\
&& + \frac{3}{8}t_0 n_B + \frac{1}{16}t_{31} n_B^{4/3} + \frac{1}{16}t_{32} n_B^{5/3} + \frac{1}{16}t_{33} n_B^{2}  \nonumber \\
&& - (1-2Y_p)^2 \Bigg[ \frac{1}{8} (t_0 + 2y_0) n_B +\frac{1}{48} (t_{31} +2 y_{31}) n_B^{4/3}   \nonumber \\
&& + \frac{1}{48} (t_{32} + 2y_{32}) n_B^{5/3} + \frac{1}{48} (t_{33} + 2y_{33}) n_B^{2} \Bigg], 
\label{eq:e0}
 \end{eqnarray}
where $m_p^*$ and $m_n^*$ are the effective masses for proton and neutron, respectively. 
The explicit expressions for $m_b^*$ ($b = p, n$) is given by 
\begin{eqnarray}
\frac{m_b^* (n_B, Y_p)}{m_N} &=& \left[ 1 + \frac{m_N}{8 \hbar^2} \left\{(3t_1+5t_2+4y_2) \pm (1-2Y_p)^2 (t_1+2y_1 - t_2 - 2y_2) \right\} n_B\right]^{-1}  \nonumber \\
&=&  \frac{1}{m_N} \left[ \frac{1}{m^*_{IS}} \pm (1-2Y_p)^2 \left(\frac{1}{m^*_{IV}} -\frac{1}{m^*_{IS}}\right) \right]^{-1}.
\label{eq:mstar}
 \end{eqnarray}
Here the upper (lower) sign corresponds to the proton (neutron), and the isoscalar nucleon effective mass $m^*_{IS}$ and isovector nucleon effective mass $m^*_{IV}$ can also be defined as 
\begin{equation}
\frac{m_{IS}^* (n_B)}{m_N} = \left[ 1 + \frac{m_N}{8 \hbar^2} (3t_1+5t_2+4y_2) n_B\right]^{-1}, 
\end{equation}
\begin{equation}
\frac{m_{IV}^* (n_B)}{m_N} = \left[ 1 + \frac{m_N}{4 \hbar^2} (2t_1+y_1 +2t_2+y_2) n_B\right]^{-1}. 
\end{equation}

The determination of specific parameters in Eq. (\ref{eq:skyrme}) follows the steps outlined below; 
For symmetric nuclear matter at $Y_p$ = 1/2, $t_0$, $t_{31}$, $t_{32}$, and $t_{33}$ are determined from the saturation properties and we assume $t_{33} = 0$ because $\alpha_n$ 
truncates at $n=2$. Here, $t_0$, $t_{31}$ and $t_{33} (=0)$ are
determined uniquely, but we cannot fix $t_{32}$ because the coefficient of the term proportioal to $n^{5/3}_B$, $\alpha_2$ in Eq.~(\ref{eq:kids}), contains $t_1$, $t_2$ and $y_2$ as well as $t_{32}$.
For pure neutron matter, 
since $t_0$, $t_{31}$ and $t_{33}$ are now known,
$y_0$, $y_{31}$ and $y_{33}$ can be also determined from the values of saturation parameters. 
On the other hand, the equation for $\alpha_2 + \beta_2$ has six parameters
$t_{32}$, $y_{32}$, $t_1$, $y_1$, $t_2$ and $y_2$.
For the $n^{5/3}_B$ terms, we have two conditions for symmetric and pure neutron matter, but there are six parameters so they cannot be determined.
In the KIDS0 model, to overcome the difficulty, we reduce the number of parameters by assuming $y_1=y_2=0$.
Furthermore, since $\alpha_2$ and $\beta_2$ are obtained prior to 
$t_{32}$, $y_{32}$, $t_1$ and $t_2$, we do not know how much portion
of $\alpha_2$ and $\beta_2$  comes from the density term ($t_{32}$ and
$y_{32}$) and from the momentum term ($t_1$ and $t_2$).
To resolve the ambiguity, we introduce a weight parameter $k$ between 0 and 1,
and evaluate the $n^{5/3}_B$ term as a weighted sum of the density term and momentum term. 
Then, $t_{32}$, $y_{32}$, $t_1$, $y_1$, $t_2$ and $y_2$ are obtained as functions of $k$, 
and only two parameters $k$ and $W_0$ are left unknown. 
In the KIDS0 model, these two parameters are fitted to the 
measured values of binding energies
and charge radii of $^{40}$Ca, $^{48}$Ca and $^{208}$Pb,
and 
we then obtain the numerical values of $t_{32}$, $y_{32}$, $t_1$, $y_1$, $t_2$ and $y_2$.
Because the fitting process proceeds in this way, 
the effective mass can be defined after the fitting to nuclear data.
We note that even though the contribution of the spin-orbit term proportional to $W_0$ is null
in the unpolarized nuclear matter, it is essential to include the term
in the fitting to reproduce the nuclear data accurately.

\begin{table}[t]
\begin{center}
\setlength{\tabcolsep}{10.0pt}
\begin{tabular}{cccccccccc}\hline
 & $t_1$ & $y_1$ & $t_2$ & $y_2$ & $t_{32}$ &  $y_{32}$ & $W_0$ & $\mu_{IS}$ & $\mu_{IV}$ \\ \hline
KIDS0 & 275.72 & 0 & $-161.51$ & 0 &  571.07 & 29485.42 & 108.36 & 0.99 & 0.82 \\ 
m*77 & 451.80 & $-400.40$ & 77.09 & $-213.05$ & $-2572.65$ & 37593.40 & 145.23 & 0.7 & 0.7 \\
m*78 & 451.92 & $-515.08$ & $-17.07$ & $-95.44$ & $-2572.65$ & 40782.86 & 148.29 & 0.7 & 0.8 \\
m*79 & 452.02 & $-603.53$ & $-91.29$ & $-2.73$ & $-2572.65$ & 43263.55 & 150.61 & 0.7 & 0.9 \\
m*71 & 452.09 & $-674.87$ & $-149.88$ & 70.45 & $-2572.65$ & 45248.10 & 152.47 & 0.7 & 1.0 \\
m*87 & 376.69 & $-265.28$ & 145.53 & $-334.84$ & $-1233.17$ & 32553.98 & 133.72 & 0.8 & 0.7 \\
m*88 & 376.81 & $-377.75$ & 48.45 & $-213.57$ & $-1233.17$ & 35743.43 &136.71 & 0.8 & 0.8 \\
m*89 & 376.90 & $-466.12$ & $-25.87$ & $-120.74$ & $-1233.17$ & 38224.12 & 139.06 & 0.8 & 0.9 \\
m*81 & 376.97 & $-537.02$ & $-85.06$ & $-46.81$ & $-1233.17$ & 40208.67 & 140.93 & 0.8 & 1.0 \\
m*91 & 318.99 & $-431.47$ & $-33.15$ & $-140.20$ & $-191.34$ & 36289.12 & 131.83 & 0.9 & 1.0 \\
m*11 & 273.01 & $-367.16$ & 34.54 & $-247.93$ & 642.12 & 33153.48 & 125.25 & 1.0 & 1.0 \\
\hline
\end{tabular}
\end{center}
\caption{Skyrme force parameters of the models for the given isoscalar and isovector effective masses at the saturation density. Units of $t_1$, $y_1$, $t_2$, $y_2$, $t_{32}$, $y_{32}$ and $W_0$ are MeV fm$^5$. Effective mass ratios are defined as $\mu_{IS} = m^*_{IS}(n_0) /m_N$ and $\mu_{IV} = m^*_{IV}(n_0)/m_N$. For simplicity, we choose $m_N$ as the mass of a neutron in numerical calculations. 
All the models have identical values of $t_0=$ -1772.04 and $y_0=$ -127.52  in MeV fm$^3$, $t_{31}=$ 12216.73 and $y_{31}=$ -11970.00 in  MeV fm$^4$, and $t_{33}=$ 0 and $y_{33}=$ -22955.30 in MeV fm$^6$.}
\label{tab1}
\end{table}

\begin{figure}[t]
\includegraphics[width=0.5\textwidth]{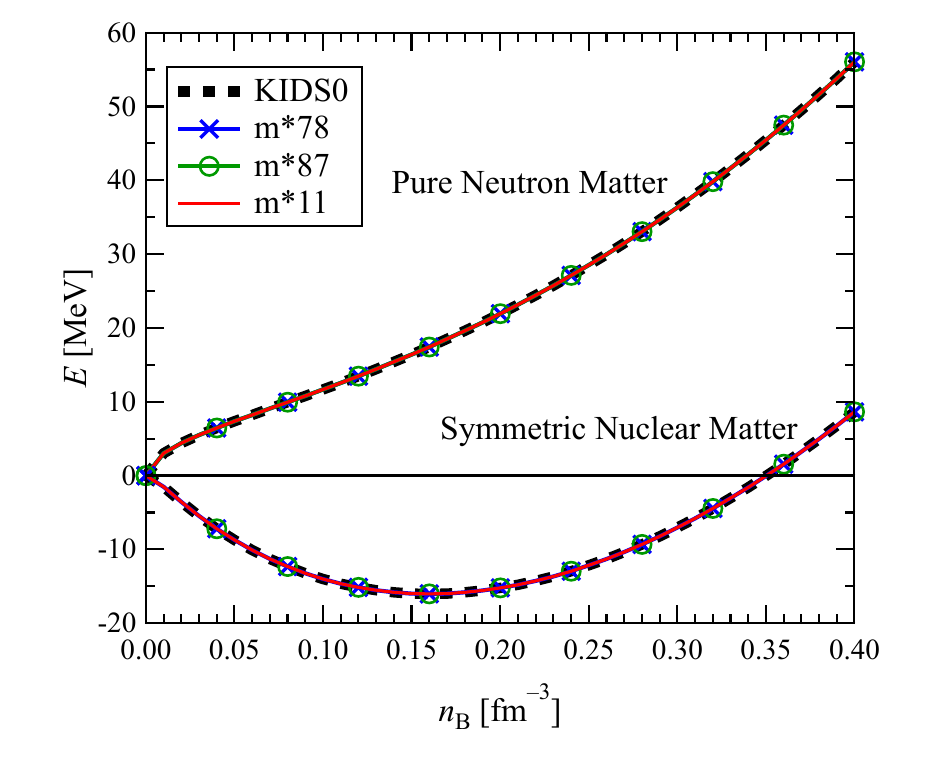}
\caption{Energy per particle for pure neutron matter (upper lines) and symmetric nuclear matter (lower lines) as a function of the density 
at $T=$ 0 with the KIDS0, KIDS0-m*78, KIDS0-m*87 and KIDS0-m*11 models.}
\label{fig:e}
\end{figure}

Table \ref{tab1} summarizes the resulting values of the parameters for the models we are considering in the work.
In the name of the model m* means the effective mass, the first number denotes $\mu_{IS} \equiv m^*_{IS} (n_0)/m_N$ and the second corresponds to $\mu_{IV} \equiv m^*_{IV} (n_0)/m_N$.
For example, m*87 means $\mu_{IS}=0.8$, $\mu_{IV}=0.7$, while
the number 1 implies 1.0.
Since the nuclear matter properties are determined prior to the fitting to nuclear properties, nuclear matter EoS is not affected by the values of the effective mass. 
Figure \ref{fig:e} shows the energy per particle for the symmetric
and pure neutron matter at zero temperature with the KIDS0, KIDS0-m*78, KIDS0-m*87 and KIDS0-m*11 models.
The figure confirms that the nuclear matter EoS at $T=0$ is not
affected by the effective mass.
Among the Skyrme force parameters, $t_1$, $y_1$, $t_2$, $y_2$, $t_{32}$, $y_{32}$ are related to the effective mass and $t_0$, $y_0$, $t_{31}$, $y_{31}$, $t_{33}$, $y_{33}$ are determined from the nuclear matter EoS and independent of the properties of finite nuclei. Therefore the four models have the same values for $t_0$, $y_0$, $t_{31}$, $y_{31}$, $t_{33}$, $y_{33}$, but differ in $t_1$, $y_1$, $t_2$, $y_2$, $t_{32}$, $y_{32}$.
Although not shown in the figure, the energies calculated using all the models in Table \ref{tab1} are difficult to distinguish from each other at $T=0$.

\begin{figure}[t]
\begin{center}
\subfloat[]{\includegraphics[width=0.5\textwidth]{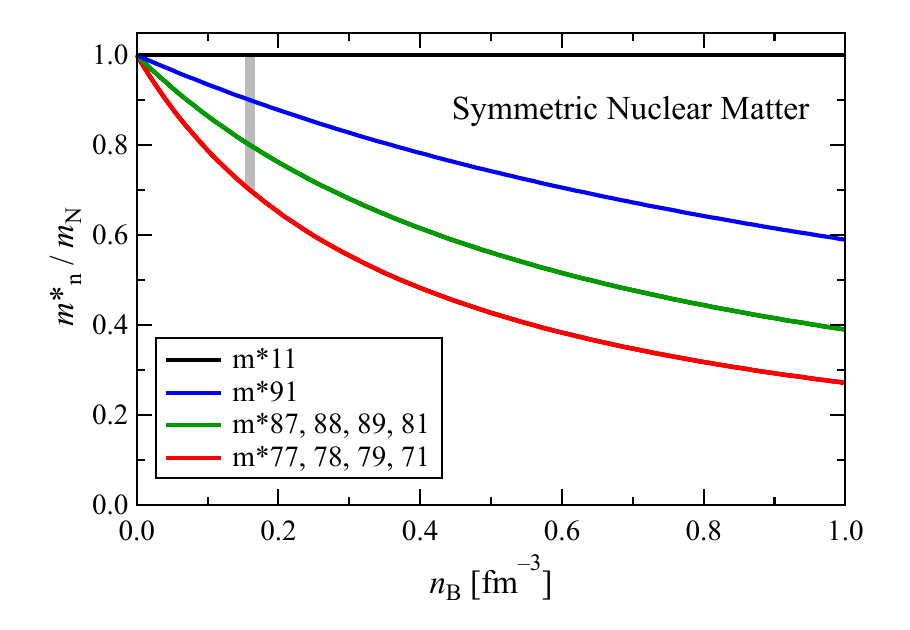}}
\subfloat[]{\includegraphics[width=0.5\textwidth]{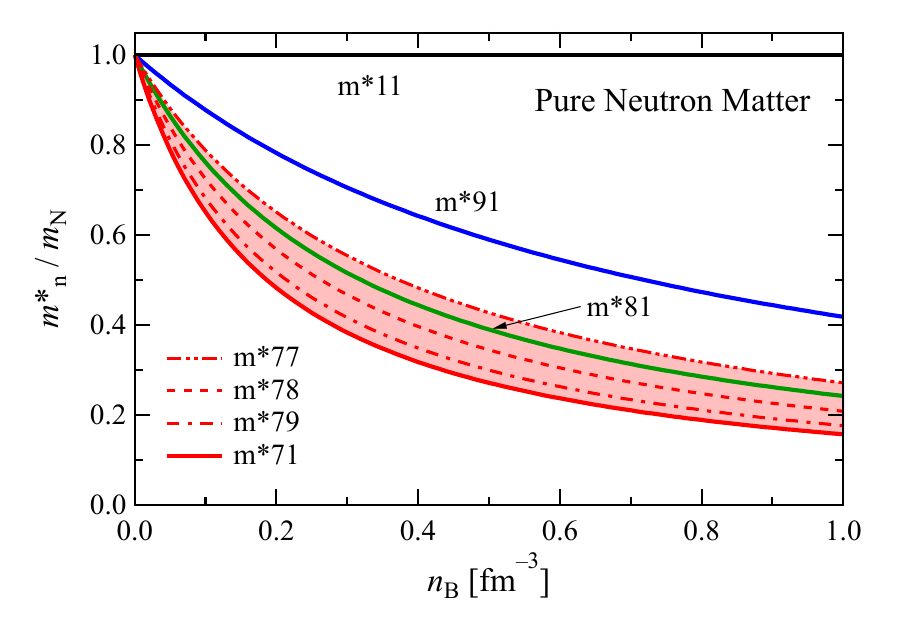}} \\
\end{center}
\caption{(a) Neutron effective mass as a function of the density for symmetric nuclear matter (b) Neutron effective mass as a function of the density for pure neutron matter.}
\label{fig:em}
\end{figure}

Figure \ref{fig:em} shows the neutron effective mass for (a) symmetric nuclear matter and (b) pure neutron matter with various KIDS models. 
In the case of symmetric nuclear matter, the effective mass of the neutron is exactly equal to that of the proton and the isoscalar effective mass, i.e., $m^*_n = m^*_p = m^*_{IS}$. 
It can be seen that $m_n^*$ decreases monotonically as $n_B$ increases. For a fixed $n_B$, $m_n^*$ decreases with $\mu_{IS}$ and is independent of the value of  $\mu_{IV}$. 
The vertical shaded region represents the empirical range of the isoscalar effective mass $(0.7-1.0)m_N$ \cite{Li:2018lpy}, and all KIDS models used in this study are within this empirical range.

In the case of pure neutron matter, $m_n^*$ decreases with $\mu_{IS}$, similar to symmetric nuclear matter, but it is observed to increase as $\mu_{IV}$ decreases. 
As a result, in the present models, when $\mu_{IS}$ is fixed at 0.7, the uncertainty in $\mu_{IV}$ relative to the neutron effective mass $m_n^*$ corresponds to the red band in Fig.~\ref{fig:em} (b). 
On the other hand, 
the effective mass of the proton $m^*_p$ completely coincides with the isovector effective mass $m^*_{IV}$  in pure neutron matter, as can be understood from Eq.~(\ref{eq:mstar}). 
Therefore, in pure neutron matter, $m^*_p$ is independent of $\mu_{IS}$ and decreases with $\mu_{IV}$.
Here, since $m^*_{IS}$ and $m^*_{IV}$ are given as functions of density with exactly the same functional form, 
the figure for $m^*_{IV}$ will be exactly the same as Fig.~\ref{fig:em} (a). In this case, 
the black line corresponds to the m*71, m*81, m*91, and m*11 models, 
the blue line corresponds to the  m*79 and m*89 models, the green line corresponds to the  m*78 and m*88 models, and the red line corresponds to the m*77 and m*87 models.

In the $T >$0 scenario, the thermodynamic quantities are expressed by the following equations. The nucleon number density $n_b$ ($b = p, n$) is given as
\begin{equation}
n_b=\frac{1}{\pi^2}\int_0^\infty f_b (k) k^2 dk,
\label{eq:nb}
\end{equation}
where the average occupation probability is 
\begin{equation}
f_b (k) = \left[1+\exp \left(\frac{\varepsilon_b(k)-{\hat \mu}_b}{k_BT} \right)\right]^{-1}.   
\end{equation}
Here, $\varepsilon_b=\hbar^2 k^2/(2m_b^*)$  and ${\hat \mu}_b$ is dertermined so as to satisfy the normalization condition in Eq.~(\ref{eq:nb}). 
It should be noted that ${\hat \mu}_b$ is different from the true chemical potential for nucleon $\mu_b$, which will later be derived from thermodynamic relations. 

The internal energy per particle $U$ is then given as
\begin{eqnarray}
U (n_B, Y_p, T) &=& \frac{\hbar^2}{2m_p^*} \frac{1}{\pi^2n_B} \int_0^\infty f_p (k) k^4 dk + \frac{\hbar^2}{2m_n^*} \frac{1}{\pi^2n_B} \int_0^\infty f_n (k) k^4 dk \nonumber \\
&& + \frac{3}{8}t_0 n_B + \frac{1}{16}t_{31} n_B^{4/3} + \frac{1}{16}t_{32} n_B^{5/3} + \frac{1}{16}t_{33} n_B^{2}  \nonumber \\
&& - (1-2Y_p)^2 \Bigg[ \frac{1}{8} (t_0 + 2y_0) n_B +\frac{1}{48} (t_{31} +2 y_{31}) n_B^{4/3}   \nonumber \\
&& + \frac{1}{48} (t_{32} + 2y_{32}) n_B^{5/3} + \frac{1}{48} (t_{33} + 2y_{33}) n_B^{2} \Bigg]. 
\label{eq:ff}
\end{eqnarray}
In Eq.(\ref{eq:ff}), the terms in the lines from 2nd to 4th ones correspond to the potential energy, which is temperature independent and is in complete agreement with the case of Eq.~(\ref{eq:e0}).
Namely, the temperature dependence is fully contained in the kinetic component.

The entropy per particle $S$ can also be expressed as
\begin{eqnarray}
S (n_B, Y_p, T) = -\frac{k_B}{\pi^2 n_B} \sum_{b = p, n} \int_0^\infty \Big[(1-f_b(k))~ \ln (1-f_b(k)) + f_b(k)~\ln f_b(k)\Big] k^2dk.    
\end{eqnarray}
Then, we obtain the free energy per particle $F(n_B, Y_p, T) = U(n_B, Y_p, T)-TS(n_B, Y_p, T)$. 
In these numerical calculations, the Fermi integrals and their inverses are computed using the subroutines provided by Fukushima \cite{F1, F2}.

The pressure $P$ and the neutron chemical potential $\mu_n$ are given by
\begin{eqnarray}
P (n_B, Y_p, T) = n_B^2 \Big(\frac{\partial F}{\partial n_B}\Big)_{Y_p,T},
\end{eqnarray}
and
\begin{eqnarray}
\mu_n (n_B, Y_p, T) = \frac{\partial}{\partial n_n} \Big(n_B F \Big)_{n_p,T} . 
\end{eqnarray}

Note that in the aforementioned finite-temperature framework, 
the nucleon effective masses are completely independent of the temperature and are given by the definition in Eq.~(\ref{eq:mstar}).

\section{Numerical Result of thermodynamic quantity}
\label{Sec:result}

In this section, we discuss the properties of various thermodynamic quantities mentioned above.

\begin{figure}[!ht]
\begin{center}
\includegraphics[width=0.5\textwidth]{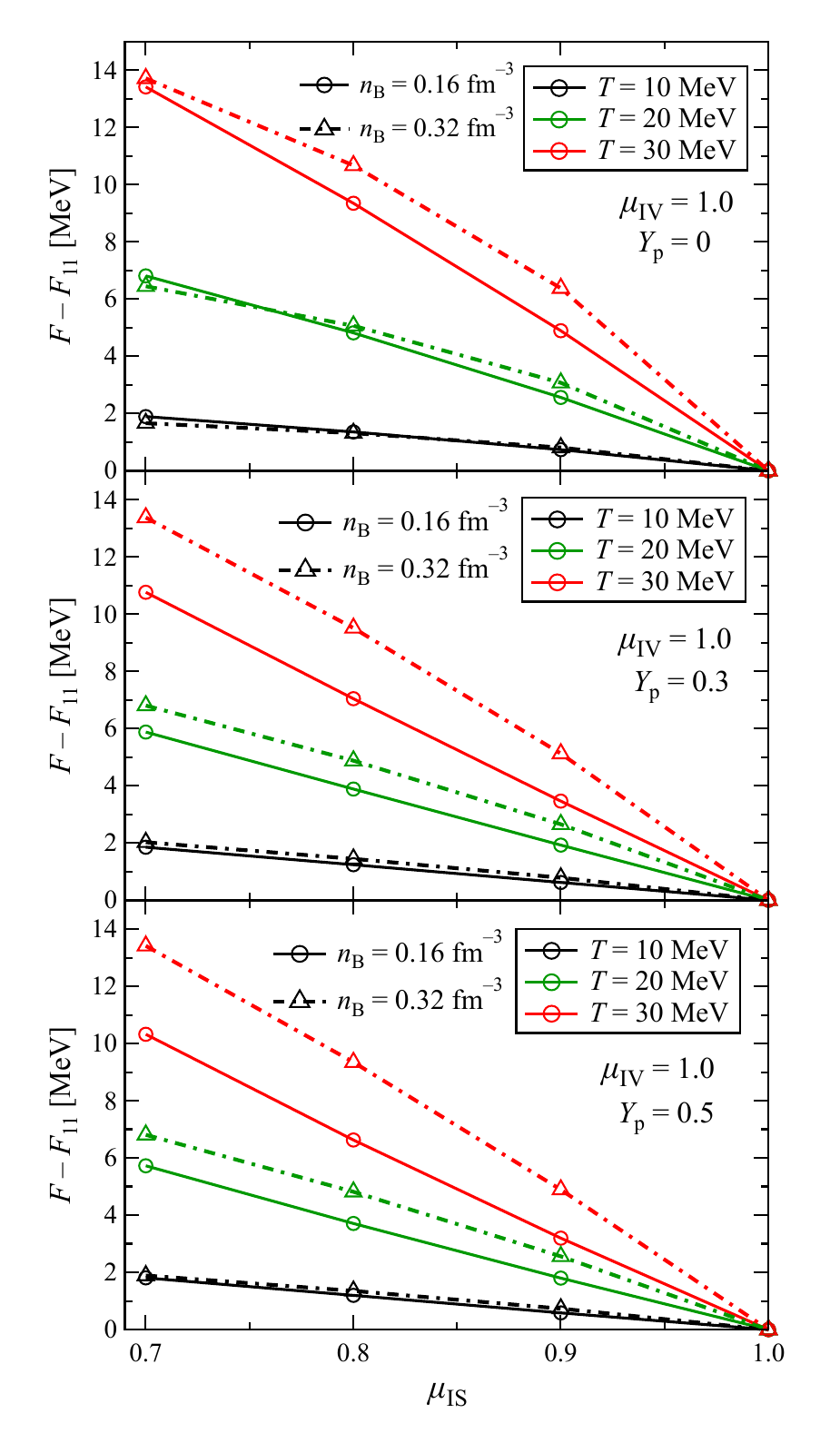}
\end{center}
\caption{Free energy residual from the values of m*11 model ($F_{11}$)
as a function of the isoscalar effective mass ratio at different temperature, density and 
proton fraction.}
\label{fig:f1}
\end{figure}


In Fig. \ref{fig:f1}, we show the difference of free energy 
with respect to the values of the m*11 model as functions of
the isoscalar effective mass ratio $\mu_{IS}$ at different temperature, density
and proton fraction.
Isovector effective mass ratio is fixed to $\mu_{IV}=1.0$.
Free energy becomes larger with small $\mu_{IS}$ and high temperature.
At $T=10$ MeV, the free energy depends weakly on the density and proton fraction, so the difference is almost identical in all the three panels.
Variation with respect to $\mu_{IS}$ becomes significant as the 
temperature increases.
At $n=2 n_0$, the result is not sensitive to temperature and proton fraction, so the lines corresponding to triangles are similar to each other.
At $n=n_0$, increase of the free energy with higher temperature is most prominent in the pure neutron matter ($Y_p=0$),
but the dependence on the temperature becomes marginal as the proton fraction increases.
Summarizing the result of 
Fig. \ref{fig:f1}, isoscalar effective mass gives an evident impact on the free energy, 
and the dependence on the isoscalar effective mass becomes stronger at high temperature and proton fraction.

\begin{figure}[t]
\begin{center}
\includegraphics[width=0.5\textwidth]{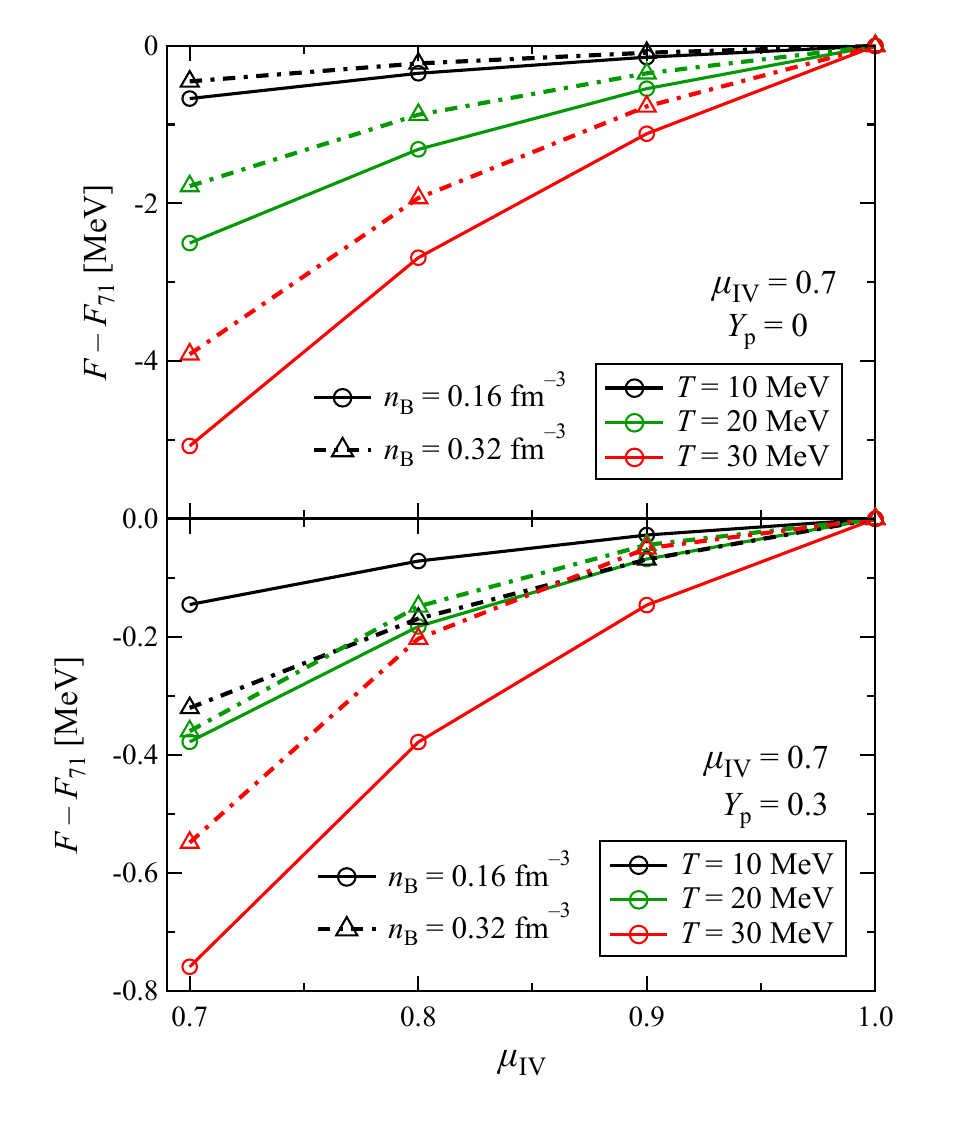}
\end{center}
\caption{Free energy residual from the values of m*71 model ($F_{71}$) as a function of the isovector effective mass ratio at different temperature, density and proton fraction.}
\label{fig:f2}
\end{figure}

Figure \ref{fig:f2} displays the free energy subtracted by the values of m*71 model 
as functions of the isovector effective mass ratio $\mu_{IV}$ with different temperature, density and proton fraction. 
Isoscalar effective mass ratio is fixed to $\mu_{IS}=0.7$.
It is notable that the magnitude of the residual ($\delta F_{IV} \equiv F- F_{71}$) depends strongly on the proton fraction.
The result is enhanced with smaller proton fraction.
$\delta F_{IV}$ is also significantly affected by both temperature and density.
Enhancement of $\delta F_{IV}$ with fewer protons becomes more
amplified at high temperature, while the dependence on the density is complicated.
Since the effect of $\mu_{IV}$ becomes comparable to that of $\mu_{IS}$
in the neutron-rich systems, both isoscalar and isovector masses must be treated carefully in the applications especially at high temperatures.

\begin{figure}[!ht]
\begin{center}
\includegraphics[width=0.5\textwidth]{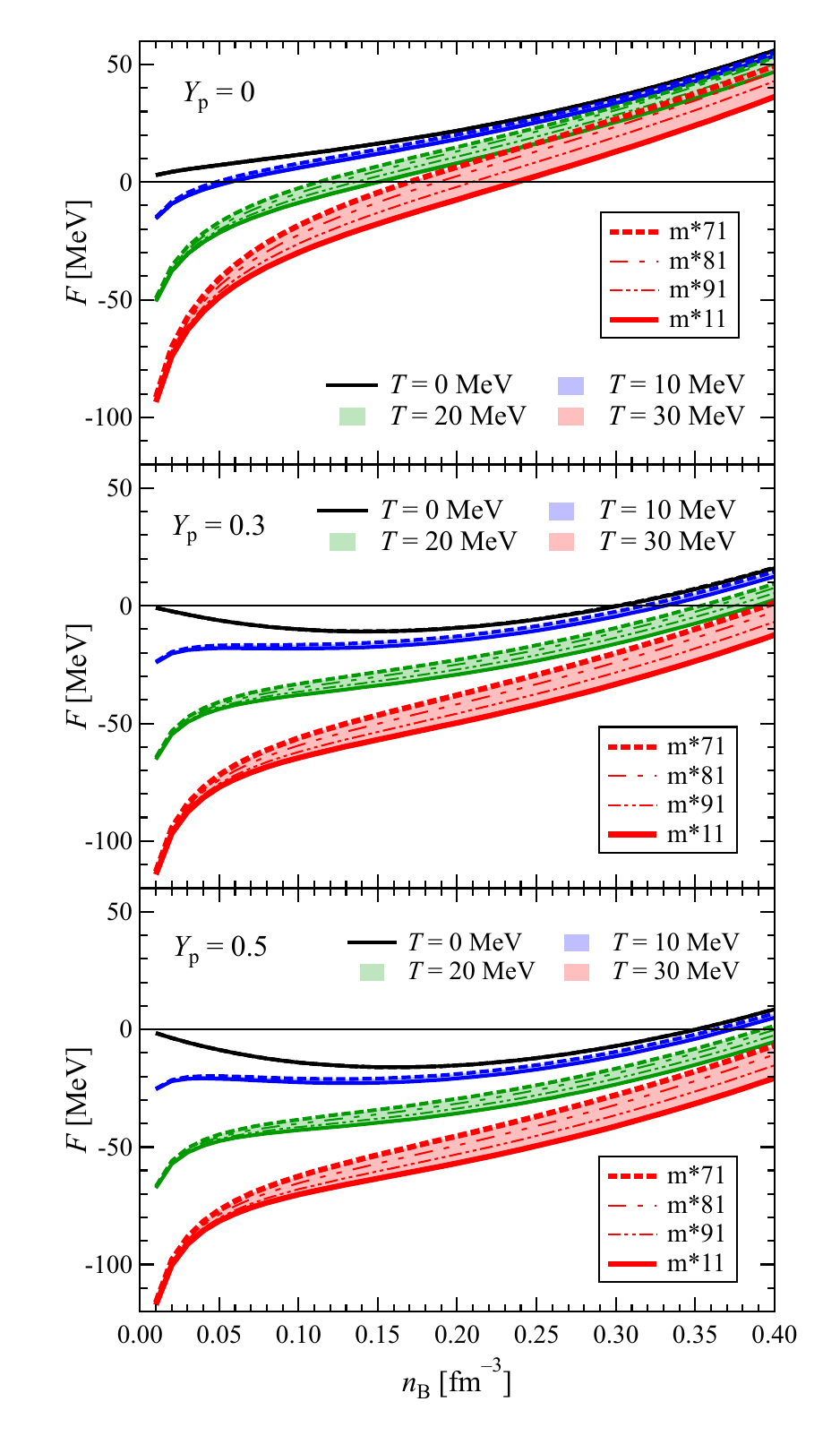}
\end{center}
\caption{Uncertainty range of the free energy as a function of density, temperature and proton fraction
with the models in Tab.~\ref{tab1}.
Uncertainty becomes enhanced at higher temperatures.
For all the temperature and proton fraction, the upper and lower bounds correspond to m*71 and m*11, respectively.}
\label{fig:f3}
\end{figure}

In Fig. \ref{fig:f3} we present the variation of free energy with respect to density at different temperature and proton fraction. 
The bands in Fig. \ref{fig:f3} indicate the uncertainty range of $F$ for the models in Tab. \ref{tab1} based on different values of $\mu_{IS}$ and $\mu_{IV}$. 
As expected, the free energy increases and becomes more positive as the matter becomes neutron-rich. 
The upper and lower bounds for each band correspond to models m*71 and m*11, respectively. 
It can be seen that uncertainty enhances as temperature increases. 
However, the width of a band at a given temperature does not differ much with different $Y_p$.
It is also worthwhile to note that regardless of temperature and proton fraction, in a given band,
the width of band increases up to the saturation density $n_0$,
and shows no significant change at higher densities.
The result shows that the uncertainty of the free energy is saturated at $n_B \gtrsim n_0$.

\begin{figure}[!ht]
\begin{center}
\subfloat[]{\includegraphics[width=0.48\textwidth]{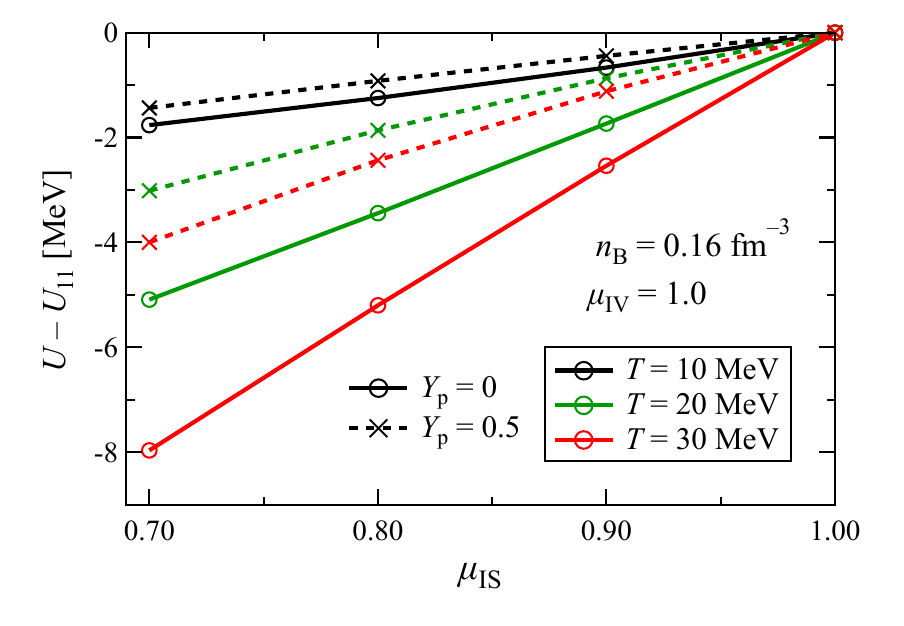}}
\subfloat[]{\includegraphics[width=0.5\textwidth]{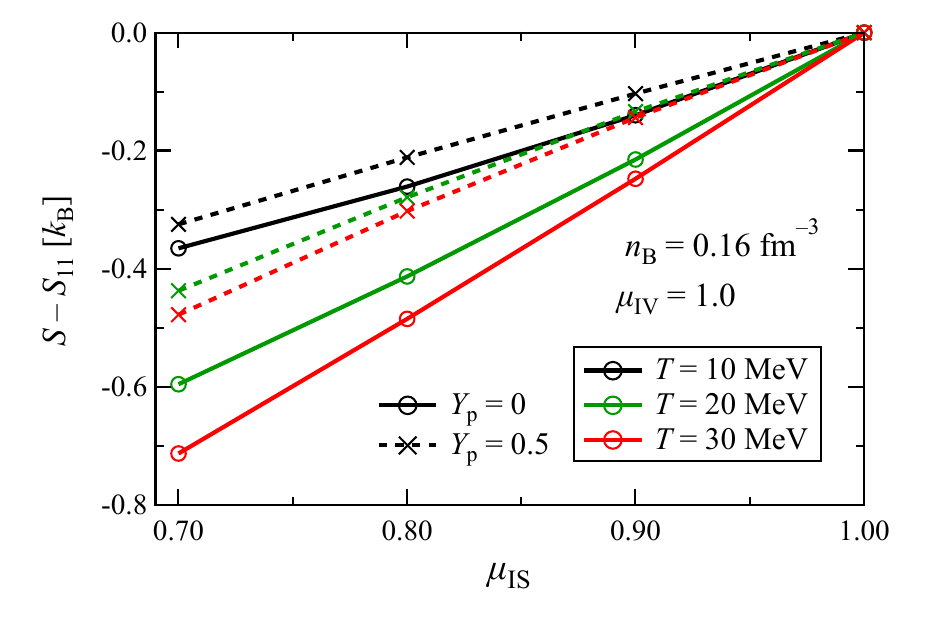}} \\
\end{center}
\caption{(a) Residual of the internal energy from $U_{11}$, (b) Residual of the entropy from $S_{11}$ as functions of $\mu_{IS}$ for symmetric and pure neutron matter with
different temperatures.}
\label{fig:u1}
\end{figure}

In Fig. \ref{fig:u1} we study the residual of the internal energy from $U_{11}$ and 
that of the entropy from $S_{11}$ with respect to $\mu_{IS}$ for pure neutron matter and symmetric matter.
Only the saturation density is considered, and the isovector 
effective mass is set to $\mu_{IV}=1.0$.
Both $U$ and $S$ increase monotonically with increasing $\mu_{IS}$.
The difference between the symmetric matter scenario and that of the pure neutron matter becomes more distinguishable with increasing temperature, 
indicating that the uncertainty due to $\mu_{IS}$ is larger at higher temperature.

\begin{figure}[t]
\centering
\subfloat[]{\includegraphics[width=0.49\textwidth]{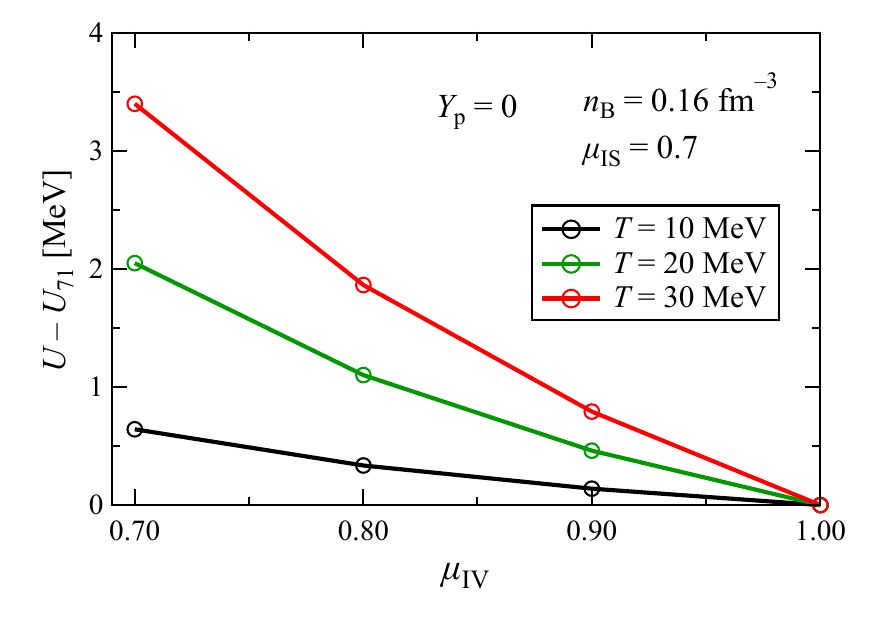}}
\subfloat[]{\includegraphics[width=0.52\textwidth]{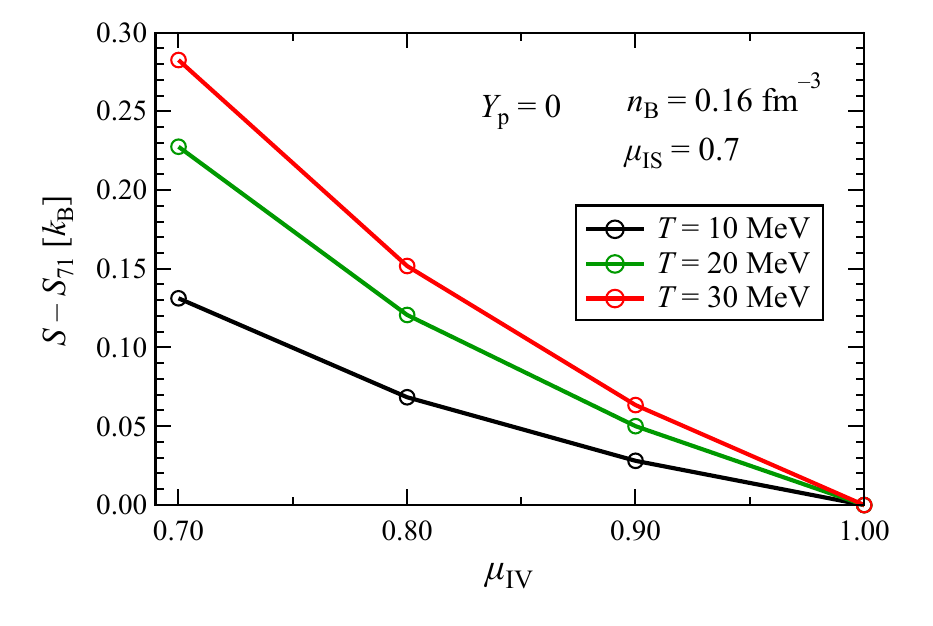}}
\caption{(a) Residual of the internal energy from $U_{71}$, (b) Residual of the entropy from $S_{71}$ as functions of $\mu_{IV}$ for pure neutron matter with different temperatures.}
\label{fig:u2}
\end{figure}

Figure \ref{fig:u2} shows the residual of the internal energy from $U_{71}$ and that of the entropy from $S_{71}$ as functions of  $\mu_{IV}$.
Since the isovector effect is maximized when the neutron-proton asymmetry becomes maximum,
we consider the pure neutron matter only.
Variations of both $U$ and $S$ follow decreasing trend. 
Dependence on temperature increases as $\mu_{IV}$ decreases.
The magnitude of uncertainty (or the value of largest difference)
at a given temperature is roughly half of those in the variation of of $\mu_{IS}$ in Fig.~\ref{fig:u1}.
For $U$ and $S$, similar to $F$, uncertainty arising from $\mu_{IS}$ is dominating, 
but the effect of $\mu_{IV}$ is also substantial.
It is shown in Fig. \ref{fig:f2} that the isovector contribution is greatly affected by the proton fraction,
so careful caution is needed when the matter is close to the pure neutron matter.

\begin{figure}[!ht]
\begin{center}
\includegraphics[width=0.51\textwidth]{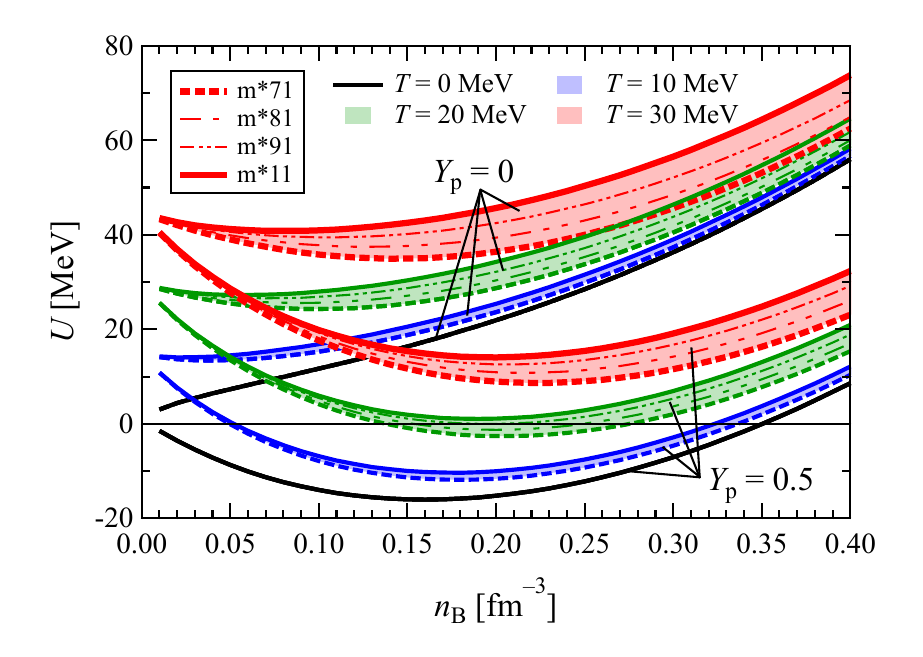}
\end{center}
\caption{Range of uncertainty for the internal energy as a function of density, temperature and proton fraction. 
For all the temperature and proton fraction, the upper and lower bounds correspond to m*11 and m*71, respectively.}
\label{fig:u3}
\end{figure}

We display the variation of internal energy with respect to density for pure neutron and symmetric matters at different temperatures in Fig.~\ref{fig:u3}. 
As expected, the value of $U$ is always positive and monotonically increasing for the pure neutron matter. 
However, in the case of symmetric nuclear matter, the value of $U$ at saturation density increases and becomes positive from negative with increasing temperature. 
In Fig.~\ref{fig:u3} we also show the uncertainty range of $U$ for different models. 
For all the temperature and proton fraction, the upper and lower bounds correspond to m*11 and m*71, respectively.
Similar to the case of $F$ seen in Fig.~\ref{fig:f3}, we find that the uncertainty in $U$ increases with temperature. 
It is also noteworthy from Fig. \ref{fig:u3} that the uncertainty in $U$ is slightly more in case of pure neutron matter than in case of symmetric nuclear matter because in the former case 
the effective mass of the nucleon $m^*_b$ depends on both the isoscalar and isovector contributions while for the latter case $m^*_b$ depends only on the isoscalar contribution. 
Width of the band (range of uncertainty) becomes wider as density increases,
but does not show recognizable change at $n_{\rm B} \gtrsim 0.3$
fm$^{-3}$.

\begin{figure}[t]
\begin{center}
\includegraphics[width=0.449\textwidth]{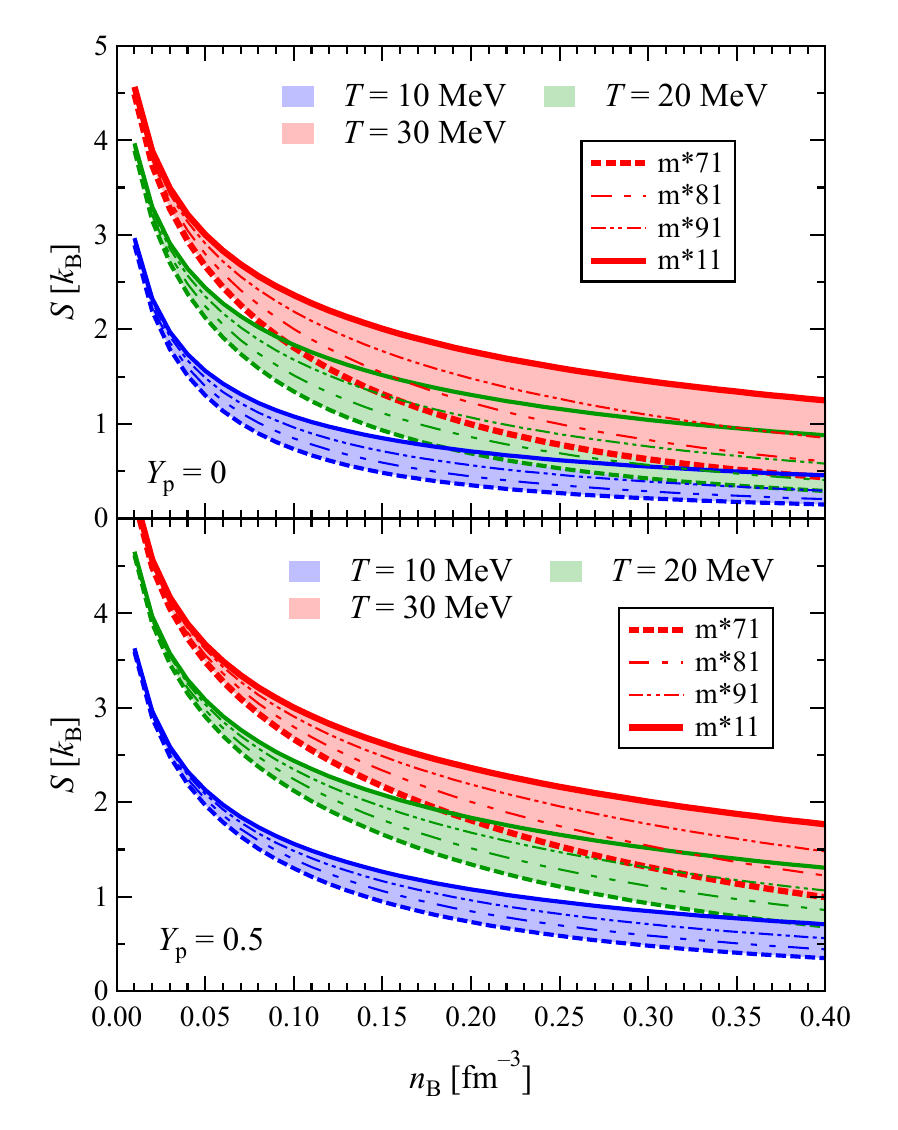}
\end{center}
\caption{Range of uncertainty for the entropy as a function of density, temperature and proton fraction. 
For all the temperature and proton fraction, the upper and lower bounds correspond to m*11 and m*71, respectively.}
\label{fig:s3}
\end{figure}

In Fig. \ref{fig:s3}, the range of entropy is shown with respect to density at different temperatures for both pure neutron and symmetric nuclear matters.
For all the temperature and proton fraction, the upper and lower bounds correspond to m*11 and m*71, respectively.
The value of entropy is larger in the symmetric matter than those in the pure neutron matter, and becomes smaller at high densities. 
Uncertainty of the entropy is magnified with increasing density and temperature.
A notable feature is that at high density and temperature, width of a band (uncertainty range) becomes comparable to the value of entropy, 
so the relative uncertainty (band width divided by the entropy value) can be more than 50\%,
which is much bigger than the relative uncertainties of $F$ and $U$.
We discuss this point in more detail at the end of this section.

\begin{figure}[t]
\begin{center}
\subfloat[]{\includegraphics[width=0.5\textwidth]{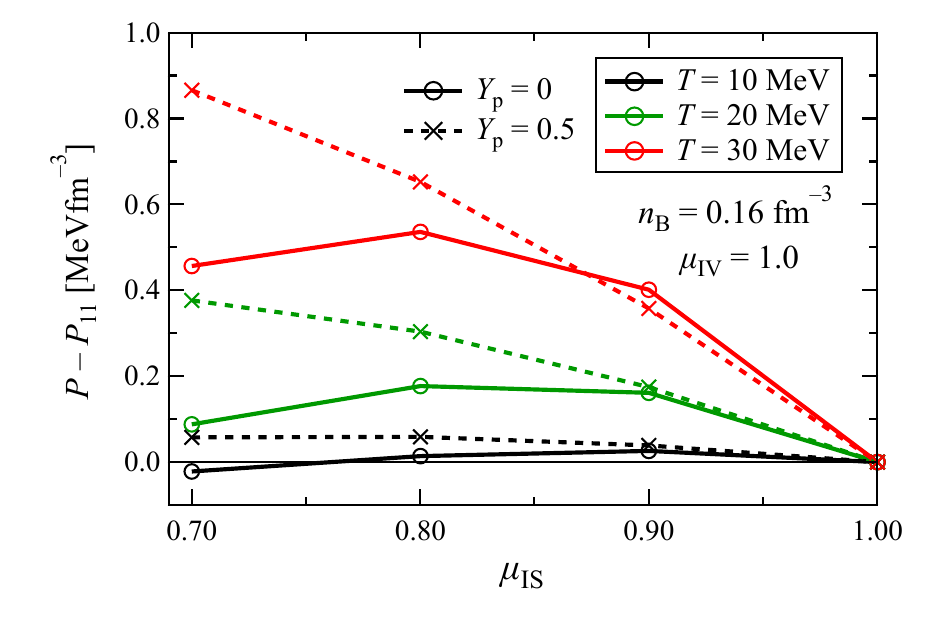}}
\subfloat[]{\includegraphics[width=0.48\textwidth]{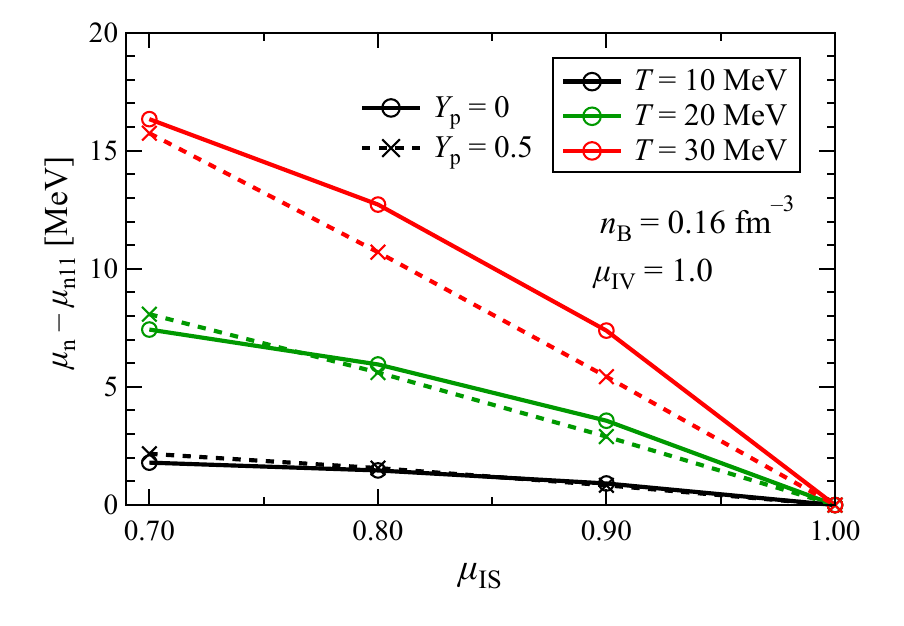}}
\end{center}
\caption{(a) Residual of the pressure from $P_{11}$ and (b) chemical potential from ${\mu_n}_{11}$ as functions of isoscalar effective mass ratio 
at different temperatures for the symmetric and pure neutron matter.}
\label{fig:p1}
\end{figure}

\begin{figure}[t]
\begin{center}
\subfloat[]{\includegraphics[width=0.5\textwidth]{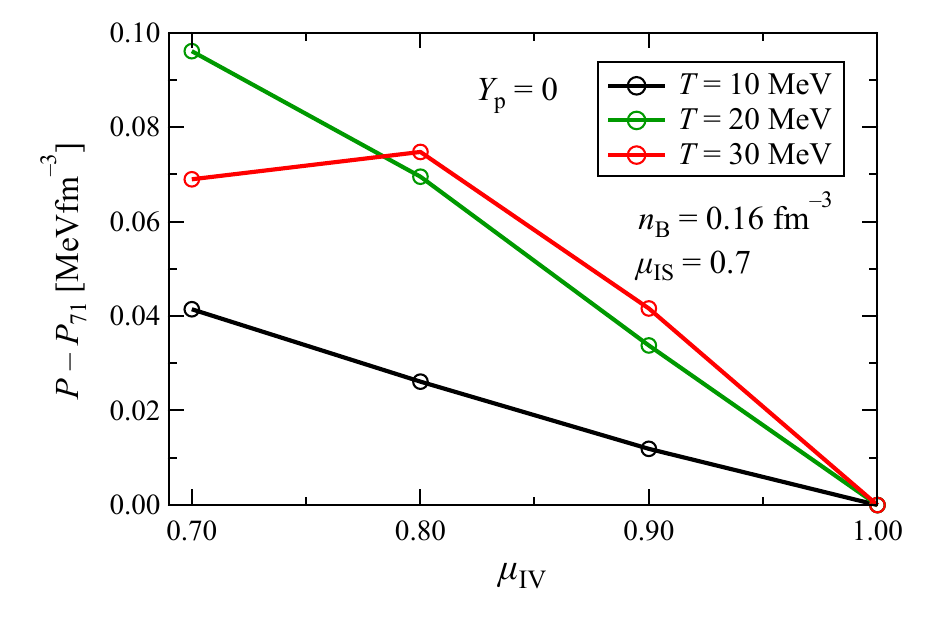}}
\subfloat[]{\includegraphics[width=0.48\textwidth]{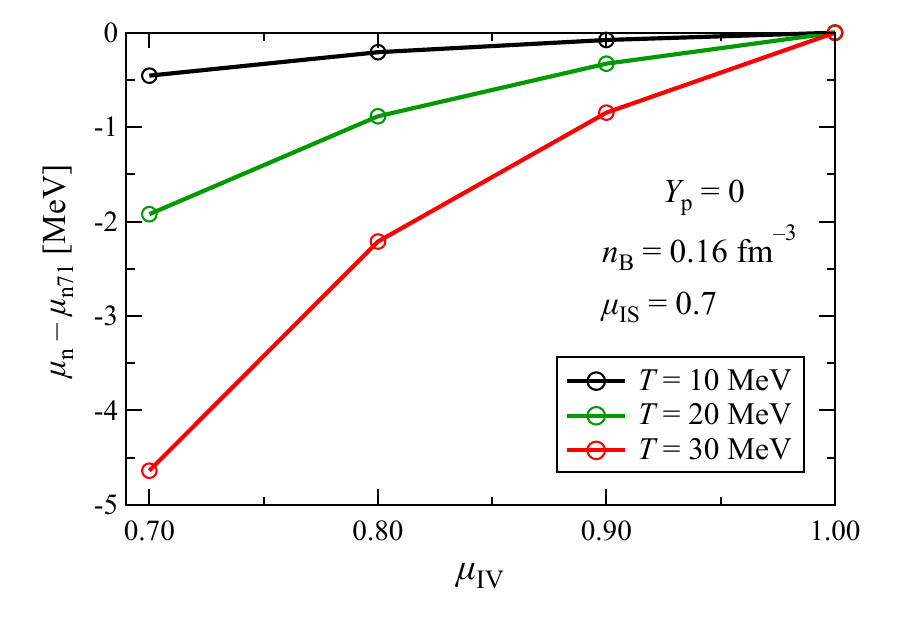}}
\end{center}
\caption{(a) Residual of the pressure from $P_{71}$ and (b) chemical potential from ${\mu_n}_{71}$ as functions of isovector effective mass ratio 
at different temperatures for the pure neutron matter.}
\label{fig:p2}
\end{figure}

Figure \ref{fig:p1} shows the variation of residual of the (a) pressure from $P_{11}$ and 
(b) neutron chemical potential from ${\mu_n}_{11}$ as functions of $\mu_{IS}$ at different temperatures for both pure neutron and symmetric nuclear matter. 
Density is fixed to the saturation density.
Figure \ref{fig:p2} also shows the same but from $P_{71}$ and ${\mu_n}_{71}$ with respect to $\mu_{IV}$. 
${\mu_n}$ shows a simple behavior, decreasing with large $\mu_{IS}$ 
but opposite with increasing $\mu_{IV}$. 
The variation of $P$ is, however, not specific or inclusive, but more complicated. 
For example, in Fig. \ref{fig:p1} the behavior of $P$ is monotonic decrease for symmetric nuclear matter, 
but it is convex for the pure neutron matter. 
This is because pressure, being the first density derivative of free energy, 
depends on the behavior of the effective mass. 
As mentioned before, for symmetric nuclear matter, the effective mass of the neutron $m_n^*$ is identical to 
that of the proton $m_p^*$ and there is no mass splitting between the neutron and the proton.
In pure neutron matter, because the isovector contribution becomes maximum,
effective masses of the neutron and the proton are distinguished, and the neutron
effective mass decreases faster than the proton effective mass with small $\mu_{IS}$.
Since the derivative of the effective mass is negative, fast decrease gives
more negative contribution to the pressure, and  this leads to relatively small
residual of $P_{71}$ compared to $P_{81}$ or $P_{91}$ in Fig. \ref{fig:p1}.
From Fig.~\ref{fig:p1} it can be seen that the sensitivity of $P$ to 
$\mu_{IS}$ is more in symmetric matter than pure neutron matter. 
From Fig.~\ref{fig:p1} we also find that the dependence of $\mu_n$ on $Y_p$ is negligible at $T=$ 10 and 20 MeV, 
and it becomes sizable at $T=$ 30 MeV, implying that the dependence of $\mu_n$ 
on $\mu_{IS}$ is magnified at higher temperatures. 
Compared to the dependence of $P$ on $\mu_{IS}$ shown in Fig.~\ref{fig:p1}, 
the sensitivity of $P$ to $\mu_{IV}$ is suppressed by an order of 10 
as seen from Fig.~\ref{fig:p2}. 
So dependence on $\mu_{IV}$ is relatively small compared to that on $\mu_{IS}$. 
Similarly, comparing Figs. \ref{fig:p1} and \ref{fig:p2} we find that the uncertainty of $\mu_n$ due to $\mu_{IV}$ is marginal compared to the uncertainty due to $\mu_{IS}$.

\begin{figure}[h]
\begin{center}
\includegraphics[width=0.5\textwidth]{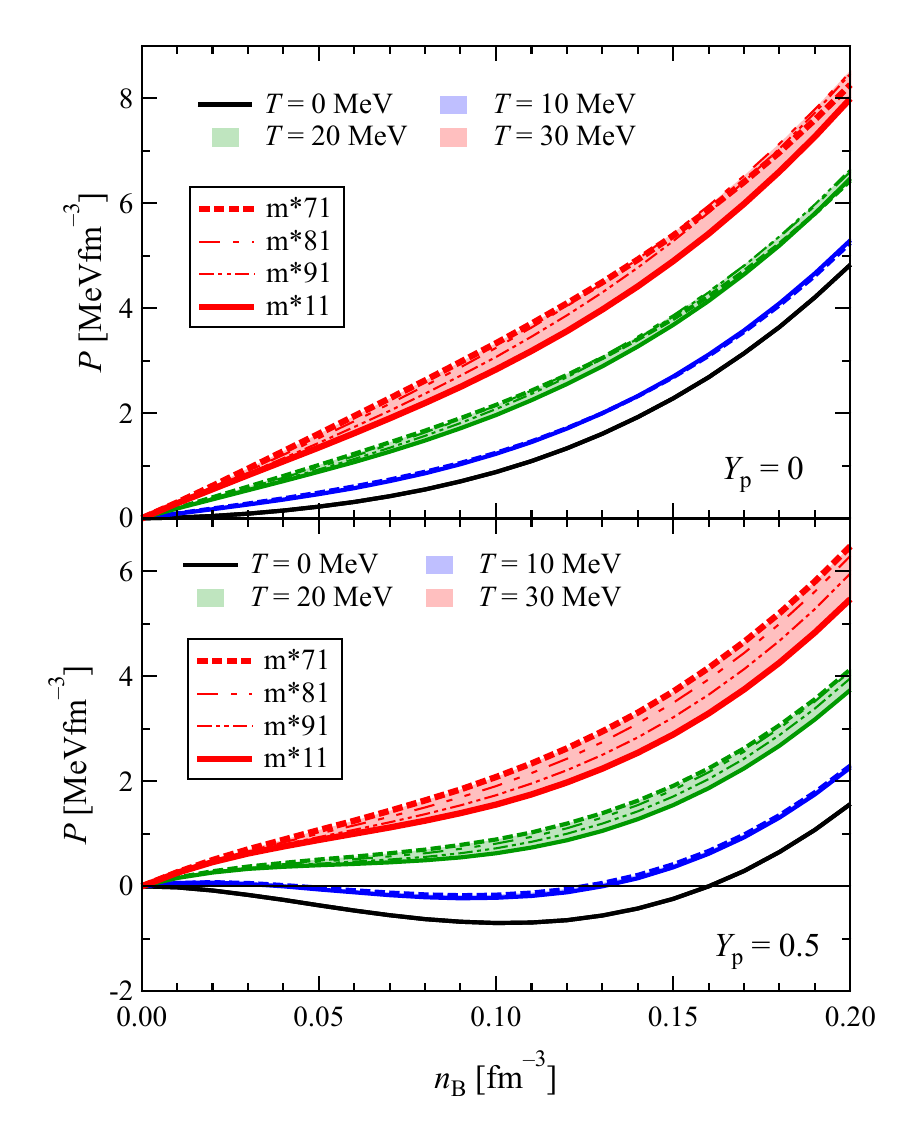}
\end{center}
\caption{Range of uncertainty for the pressure as a function of density, temperature and proton fraction.}
\label{fig:p3}
\end{figure}

\begin{figure}[t]
\begin{center}
\includegraphics[width=0.52\textwidth]{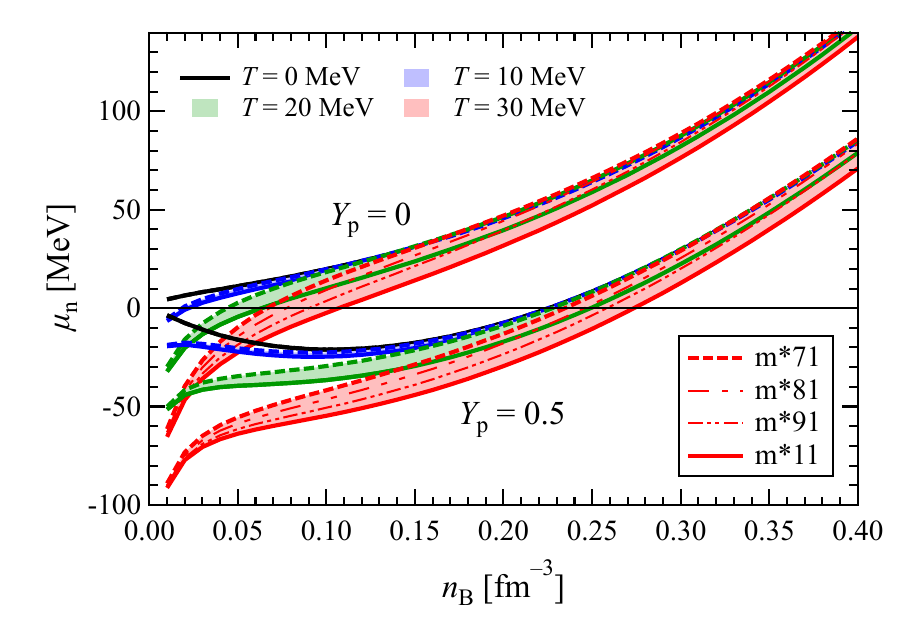}
\end{center}
\caption{Range of uncertainty for the chemical potential as a function of density, temperature and proton fraction.}
\label{fig:mu3}
\end{figure}

In Figs.~\ref{fig:p3} and \ref{fig:mu3} we depict the density dependence of
pressure and neutron chemical potential, respectively for pure neutron and symmetric nuclear matter at different temperatures. 
Both $P$ and $\mu_n$ increase with density at all densities in the pure neutron matter,
and at $n_{\rm B} \gtrsim 0.1\, {\rm fm}^{-3}$ for the symmetric nuclear matter.
Similar to the internal energy, pressure is always positive in the pure neutron matter. 
In the case of symmetric nuclear matter, the value of $P$ at minima increases and
becomes positive after reaching the critical temperature,
which is about slightly above 10 MeV.
Similar to the other thermodynamical quantities like $F$, $U$ and $S$, the uncertainty in $P$ and $\mu_n$ also enhances with increasing temperature. 
At $T=$ 10 MeV, the width of the band is very narrow, but it becomes substantial at higher temperature. 
While the width of band for $Y_p=0$ is broader than that of $Y_p=0.5$ for $F$, $U$ and $S$,
symmetric nuclear matter has a larger uncertainty than pure neutron matter for $P$.
The widths of the bands increase up to saturation density, and after that it seldom changes. 
This behavior is similar to $F$ and $U$, and opposite to $S$ for which width of the band increases as $n_B$ increases.

In the case of $P$ shown in Fig.~\ref{fig:p3}, the upper and lower bounds correspond to m*71 and m*11, respectively, for all the temperatures in symmetric nuclear matter. 
On the other hand, in pure neutron matter, the lower limit corresponds to the m*11 at all the temperatures in this figure, 
while the model that sets the upper limit varies with density. 
For $P$ at $T =$ 30 MeV, the m*71 model provides the upper bound in the low-density region $n_{\rm B} \lesssim 0.12\ {\rm fm}^{-3}$ 
but at higher densities, it is replaced by other models. 
Furtheremore, at densities even higher than those shown in Fig.~\ref{fig:p3}, 
it is also found that, in both symmetric nuclear matter and pure neutron matter, the models providing both the upper and lower limits of $P$ change with density.
In the case of $\mu_n$, it does not exhibit as complex model dependence as $P$. 
For symmetric nuclear matter, the m*71 always provides the upper limit, while the m*11 gives the lower limit at all densities and temperatures. 
For pure neutron matter, the lower limit is consistently determined by the m*11, while the model providing the upper limit changes from m*71 to another model at high densities.

From the above discussion, we notice that in case of most of the thermodynamic quantities, the isoscalar effective mass has larger effects than the isovector effective mass. 
The reason for this can be explained as follows; 
In the case of symmetric nuclear matter, as shown in Eq.~(\ref{eq:mstar}), 
both the proton and neutron effective masses coincide with the isoscalar effective mass ($m^*_n = m^*_p = m^*_{IS}$). 
Therefore, changes in $\mu_{IS}$ directly affect thermodynamic quantities through the nucleon effective masses, which do not depend on $\mu_{IV}$ at all.

In contrast, in neutron-rich matter, since $(1 - 2Y_p) \sim 1$, 
Eq.~(\ref{eq:mstar}) indicates that $m_p^*$ is primarily determined by $m_{IV}^*$, 
while $m_n^*$ is influenced by both $m_{IS}^*$ and $m_{IV}^*$, with a relative weighting of 2:1. 
Here, the number of protons is small in neutron-rich matter, so the contribution from the protons becomes relatively insignificant, 
and the behavior of neutron occupation probability thus dominates the thermodynamic quantities. 
Consequently, the effect of $\mu_{IS}$, which influences $m_n^*$ with twice the weight of $\mu_{IV}$, remains significant even in neutron-rich matter.

\begin{figure}[t]
\begin{center}
\includegraphics[width=1.0\textwidth]{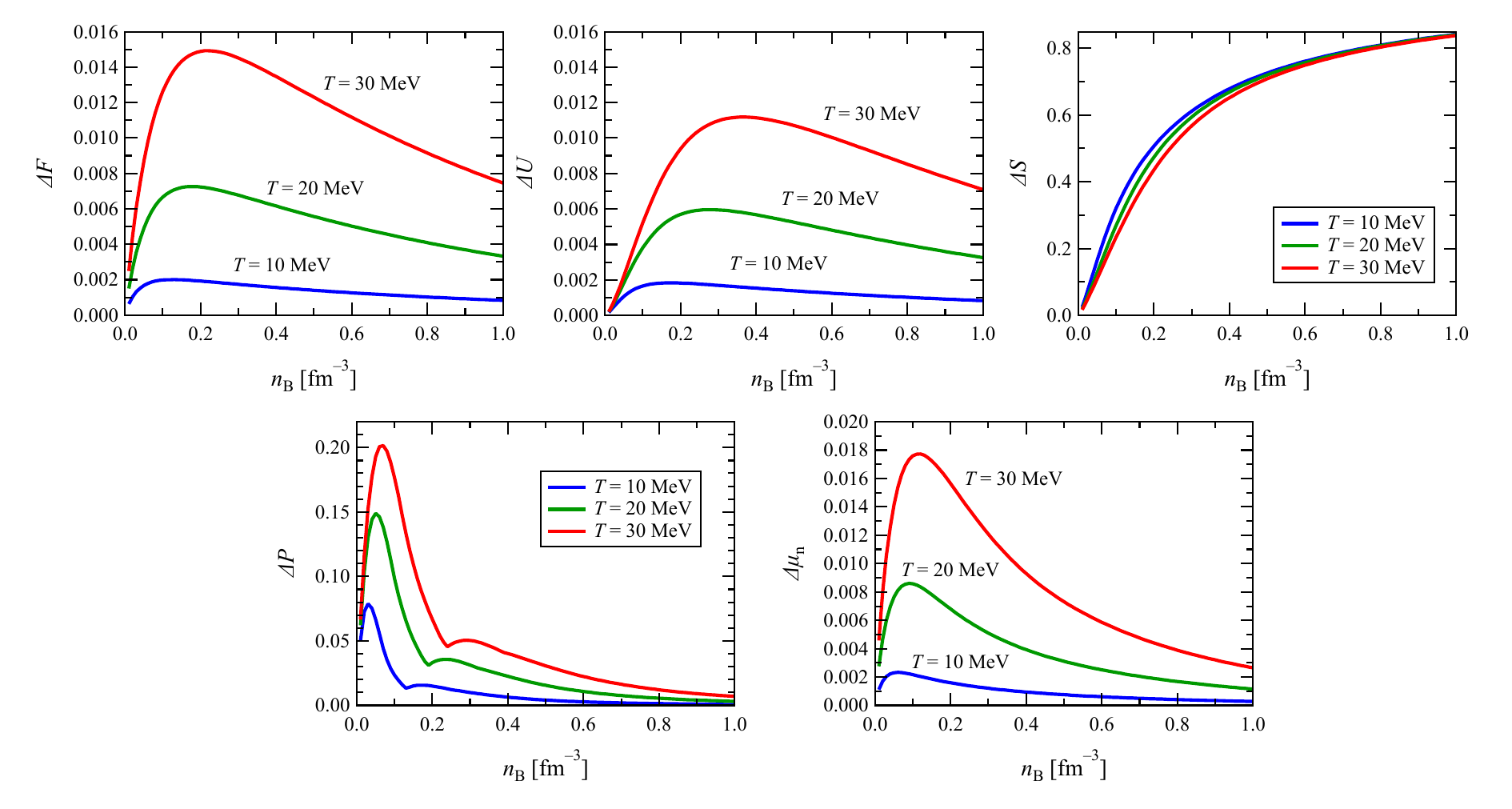}
\end{center}
\caption{Maximum uncertainty of thermodynamical quantities normalized by the quantities at $\mu_{IS}=\mu_{IV}=1.0$.}
\label{fig:err}
\end{figure}

Up to now we have seen the values of the thermodynamic quantities and their uncertainties arising from the effective mass. 
We find that the uncertainty originating from $\mu_{IS}$ is dominant and the role of $\mu_{IV}$ suppressed by a factor from 3 to 10. 
The uncertainty gets magnified with temperature, and the sensitivity to the proton fraction is also easily distinguished. 
Another way which could be more comprehensive to understand the dependence on the effective mass is 
the relative uncertainty, i.e. the maximum
difference divided by a reference value. 
For instance, entropy has a range $(1.5-2.0)k_B$ at $n_0$ in the pure neutron matter, 
and so its relative uncertainty $\Delta S \sim 0.5/2.0 - 0.5/1.5$ takes a range 25--33\%. 
For the pressure, however, the relative uncertainty is below 10\%, so the relative uncertainties are strongly dependent on the observables. 
In Fig. \ref{fig:err} we show the dependence of normalized maximum uncertainty on density. 
The normalization factors for $\Delta F$, $\Delta U$, and $\Delta \mu_n$ include the averaged nucleon mass $M = Y_p m_p + (1-Y_p) m_n$, 
so they are ($F_{11}$ + $M$), ($U_{11}$ + $M$), and (${\mu_n}_{11}$ + $m_n$), respectively,
while $\Delta S$ and $\Delta P$ do not involve $M$ in the definition 
of the relative uncertainty. 
For example, $\Delta F = \frac{|F-F_{11}|}{F_{11} + M}$ while $\Delta S = \frac{|S-S_{11}|}{S_{11}}$. 

The relative uncertainty of $F$, $U$, $P$ and $\mu_n$ increases very quickly 
as temperature increases.
Another interesting result is that at a given temperature,
the relative uncertainty reaches a maximum at a certain density,
and above the density it decreases as density increases for $F$, $U$, $P$ and $\mu_n$.
However, entropy shows behavior completely different from $F$, $U$, $P$ and $\mu_n$.
Dependence on the temperature is scarcely distinguished,
but the relative uncertain increases monotonically with density.
Entropy also exhibits unique behavior in terms of the magnitude of the relative uncertainty.
Maximum uncertainty at $T=30$ MeV does not exceed 2~\% for $F$, $U$ and $\mu_n$,
and it is about 20~\% for $P$ at most.
For $S$, however, it is about 40~\% at $n_0$ and approaches 60~\% at $2n_0$.
If an observable is considered, uncertainty due to $S$
may be a dominant source of the uncertainty for the observable,
and pressure can also have a significant effect.
On the other hand, since the uncertainty due to the effective mass is well below
2~\% for $F$, $U$ and $\mu_n$, their uncertainties may not play a critical role
to the uncertainty of an observable.
To have a sense on the effect of the uncertainty due to the effective mass,
we consider the proto-neutron star in the following section.

\section{Proto-Neutron Star}
\label{Sec:PNS}
In this section, we apply the KIDS model at finite temperature to investigate the properties of proto-neutron stars for different isoscalar effective masses. 
A proto-neutron star is a hot, lepton-rich remnant formed immediately after the gravitational collapse of a massive star in a core-collapse supernova, primarily due to neutrino trapping. 
Over tens of seconds, it undergoes deleptonization as neutrinos escape, eventually evolving into a cold neutron star or collapsing into a black hole.
To replicate such conditions, we consider charge-neutral, $\beta$-stable matter composed of nucleons, leptons, and photons, with a fixed lepton fraction per nucleon $Y_l = 0.3$. 
The lepton component consists of electrons, positrons, electron-type neutrinos, and electron-type antineutrinos, treated as relativistic non-interacting Fermi gases. 
The contribution of photons is incorporated via the Stefan-Boltzmann law. 
The thermodynamic quantities of proto-neutron star matter are computed along adiabats with a fixed total thermodynamic entropy per baryon, $S_{PNS}$, which includes contributions from nucleons, leptons, and photons. 
This is because a proto-neutron star evolves approximately along isentropic trajectories, where the total entropy per baryon $S_{PNS}$ remains nearly constant in each region of the star due to the high neutrino opacity at early times. In contrast, temperature varies significantly with density and position within the star, making it less suitable as a control parameter \cite{Prakash:1996xs}.
Since we focus on the influence of effective masses of uniform nuclear matter, we supplement the non-uniform EoS at low densities by the TNTYST EoS \cite{TNTYST}.

\begin{figure}[h]
\begin{center}
\subfloat[]{\includegraphics[width=0.47\textwidth]{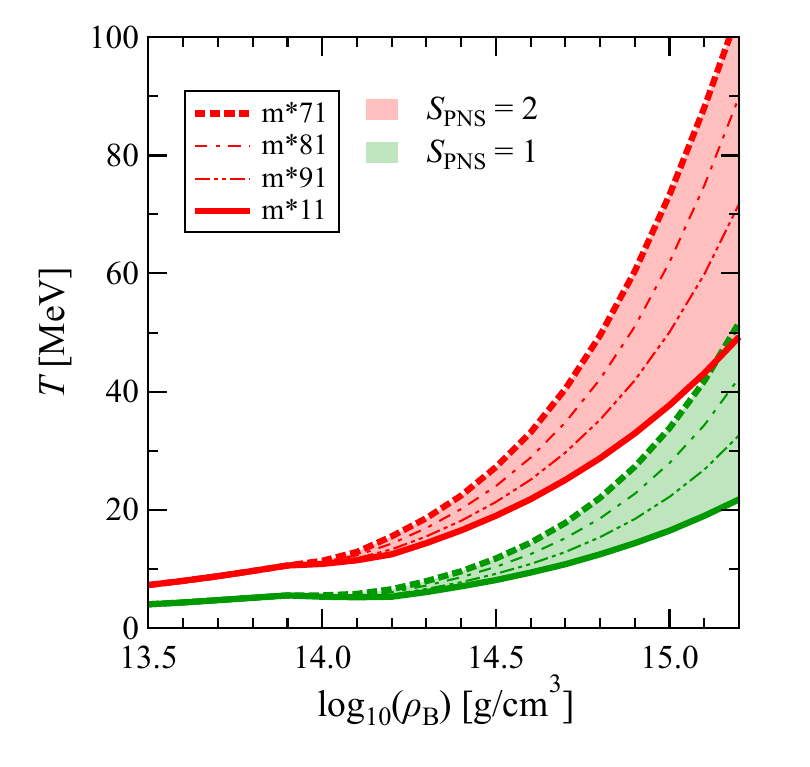}}
\subfloat[]{\includegraphics[width=0.48\textwidth]{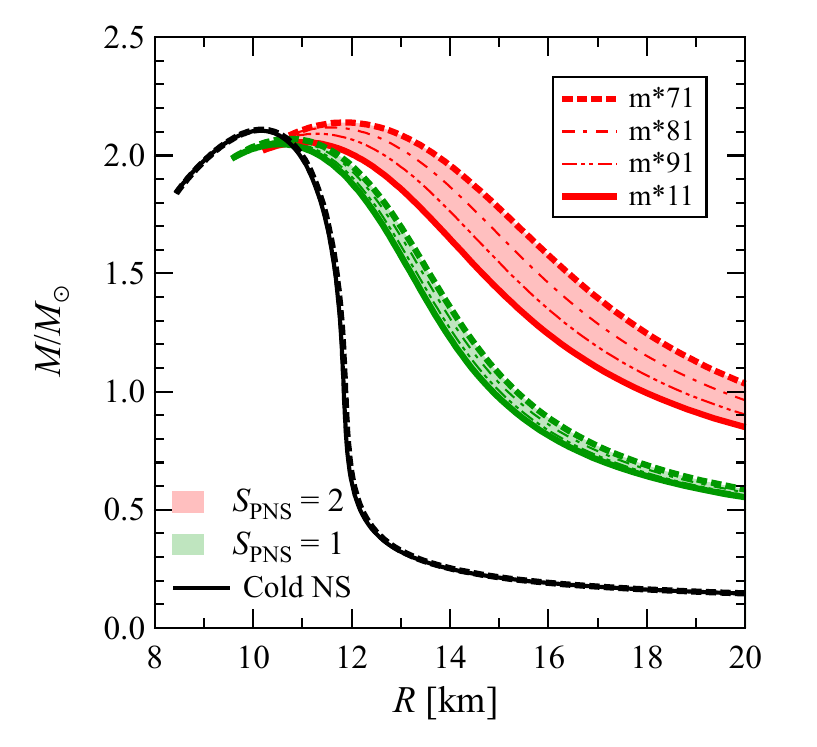}}
\end{center}
\caption{(a) Temperature profile and (b) Mass-radius relation
of the proto-neutron star for $\mu_{IS}=$ 0.7, 0.8, 0.9, and 1.0 at $S$ = 1 and 2.}
\label{fig:pns}
\end{figure}

In Fig. \ref{fig:pns} (a) we show the temperature profile in the proto-neutron star at the total entropy per baryon $S_{PNS}$ = 1 and 2 
as a function of the baryon mass density $\rho_B$ for the isoscalar effective masses $\mu_{IS}=$ 0.7, 0.8, 0.9, and 1.0.
Isovector effective mass is fixed to $\mu_{IV}=1.0$.
Noticeable difference begins at $\rho_B=$ 10$^{14}$ g/cm$^3$ because the TNTYST EoS is adopted at $\rho_B<$ 10$^{14}$ g/cm$^3$. 
The red (green) band in Fig.~\ref{fig:pns} corresponds to the uncertainty range of the temeperature profile in proto-neutron star at $S_{PNS}$ = 2 (1) 
for different models shown in Tab. \ref{tab1}. 
The range becomes enlarged as density increases. 
In the case of $S_{PNS}$ = 2 at $n_B=$ 10$^{15}$ g/cm$^3$, temperatures are about 40, 50, 60, and 70 MeV for
$\mu_{IS}=$1.0, 0.9, 0.8 and 0.7, respectively.
Temperature is sensitive to the isoscalar effective mass,
and tends to increase with lighter $m^*_{IS}$ value.
In the preceding results, it is shown that the internal energy decreases and pressure increases as $\mu_{IS}$ decreases.
Therefore, one can expect that the EoS becomes stiff at small $\mu_{IS}$ values.
Stiffness of EoS with respect to the isoscalar effective mass can be understood from Fig. \ref{fig:pns} (b), 
where we display the mass-radius relation of the proto-neutron star with the EoS obtained in Fig.~\ref{fig:pns} (a). 
For comparison, we also show the mass-radius relation of the cold neutron star with m*11 and m*71 models, 
which represent the neutrino-untrapped $\beta$-stable matter at $T=0$.
In the case of cold neutron star, the results obtained with all the KIDS models are difficult to distinguish from each other.  
In Fig.~\ref{fig:pns} (b), it can also be seen that the radius of the canonical mass star is $R_{1.4}= 11.8$ km for cold neutron star.
On the other hand, for the proto-neutron star at $S_{PNS}$ = 2, $R_{1.4}\simeq$ 17.0, 16.5, 16.0 and 15.0 km 
for $\mu_{IS}=$ 0.7, 0.8, 0.9, and 1.0, respectively,
so the result indicates significant swelling of the radius at finite temperature.
This is consistent with the results in \cite{Yu12,Sen21}. 
In general, radius becomes large with stiff EoS, so the mass-radius relation of the proto-neutron star demonstrates
close correlation between the effective mass and stiffness of EoS at finite temperature. 

In contrast to the case of cold neutron stars, where muon mixing is considered, 
muons are not included in the proto-neutron star model. 
This is because no muons are present when neutrinos become trapped and the total muon lepton number can be considered to be zero.
As the proto-neutron star evolves and the density increases, muons can be thermally produced and contribute to matter properties, 
but even then, muons play only a minor role \cite{AA2010}. Moreover, we note that many simulations of core-collapse supernovae also neglect muons in their calculations \cite{KS4,SNreview1,SNreview2}.

\section{Summary}
\label{Sec:Sum}
We examined the properties of nuclear matter at finite density and temperature, particularly emphasizing the role played by the effective mass in this context. 
For the purpose of this study, the KIDS model is adopted. 
The free energy is found to be larger (smaller) with smaller values isoscalar (isovector) effective mass, the effects being more pronounced at higher temperature. 
The internal energy and the entropy increase (decrease) with increasing isoscalar (isovector) effective mass, the magnitude of uncertainty due to isovector effective mass is about half of that of isoscalar effective mass. 
The sensitivity of isovector effective mass to the pressure residual is relatively quite small compared to that of isoscalar effective mass. Similarly, 
the uncertainty due to isovector effective mass on the  residual of chemical potential is marginal compared to that of isoscalar effective mass. 
The maximum uncertainties of free energy, internal energy, and chemical potential show that these thermodynamical quantities are weakly dependent on
the effective mass. 
However, the entropy is quite strongly affected by effective mass but weakly affected by temperature. Both effective mass and temperature have moderate effect on pressure.

We also extend our work to investigate the properties of proto-neutron stars considering the neutrino trapped case with isentropic matter. We found that smaller isoscalar effective mass makes the EoS stiffer, which results in larger values of $R_{1.4}$.

\section*{Acknowledgments}
This work was supported by the National Research Foundation of Korea (NRF) Grant
Nos. 2018R1A5A1025563 and 2023R1A2C1003177. The research of author H.T. was supported in part
by JSPS KAKENHI Grant Numbers JP21K13924 and JP21H01088. This research was funded by the
National Research Foundation of Korea (NRF) Grant No. 2023R1A2C1003177.
We would also like to express our special thanks to K.~Sumiyoshi and M.~Takano for valuable discussions and comments.

\end{document}